\documentclass[pra,twocolumn,showpacs,preprintnumbers,floatfix]{revtex4-1}
\usepackage{graphicx}
\usepackage{dcolumn}
\usepackage{bm}
\usepackage{latexsym,epsfig}
\usepackage{graphicx}
\usepackage{verbatim}
\usepackage{comment}
\usepackage{amsmath}
\usepackage{stmaryrd}
\usepackage{color}
\usepackage{float}
\usepackage{microtype}

\newcommand{\beq}{\begin{equation}}
\newcommand{\eeq}{\end{equation}}
\newcommand{\bea}{\begin{eqnarray}}
\newcommand{\eea}{\end{eqnarray}}
\newcommand{\ben}{\begin{eqnarray*}}
\newcommand{\een}{\end{eqnarray*}}
\newcommand{\bfig}{\begin{figure}}
\newcommand{\efig}{\end{figure}}

\begin{document}
\title{Permanent Electric Dipole Moment of Strontium Monofluoride as a Test of the Accuracy of a Relativistic Coupled Cluster Method}
\author{V S Prasannaa$^1$, M Abe$^2$ $^3$, B P Das$^1$}
\affiliation{$^1$Indian Institute of Astrophysics, II Block, Koramangala, Bangalore-560 034, India}
\affiliation{$^2$Tokyo Metropolitan University, 1-1, Minami-Osawa, Hachioji-city, Tokyo 192-0397, Japan}
\affiliation{$^3$JST, CREST, 4-1-8 Honcho, Kawaguchi, Saitama 332-0012, Japan}

\date{\today}

\begin{abstract}
The permanent electric dipole moment of the $X^2\Sigma^+$ electronic ground state of the strontium monofluoride 
molecule is calculated using a relativistic coupled cluster method. Our result is compared with those of other calculations   
and that of experiment. Individual contributions arising from different
physical effects are presented. The result obtained suggests that the relativistic coupled cluster method 
used in the present work is capable of yielding
accurate results for the permanent electric dipole moments of molecules for which relativistic effects cannot be ignored.

\end{abstract}


\maketitle

\section{Introduction}

The relativistic coupled cluster (RCC) method is considered to be the current gold standard of many-body 
theory of heavy atoms and diatomic molecules~\cite{Eliav}~\cite{Natarajhs}.
This method has been used extensively to calculate the properties of various 
atoms, but its applications to the properties of molecules other than their energies have been very limited to date. 

The effective electric field in a molecule is required for obtaining an upper limit on the 
electric dipole moment of the electron, which can provide insights into physics beyond 
the Standard Model. Recently, Abe et al applied a RCC method to compute the effective electric 
field experienced by an electron in the YbF molecule~\cite{Das}. 
In addition, they had also obtained the permanent electric dipole moment (pdm) and the 
magnetic dipole hyperfine coupling constant to assess the accuracy of their calculations. 
The purpose of this paper is to test the accuracy of a single reference RCC method by using it 
to calculate the pdm of the strontium monofluoride molecule (SrF), for which the interplay of relativistic
and correlation effects is important.

Sr is a moderately heavy atom, and as a result of this, the relativistic effects are more pronounced in SrF, compare to  
a fluoride of a lighter alkaline earth atom like calcium. We have chosen an alkaline earth monofluoride 
because it contains only a single valence electron, making it a suitable candidate to 
test a single reference coupled cluster method. Furthermore, a high precision measurement of the 
SrF pdm is available for comparison~\cite{Ernstt}.

SrF is the only molecule that has been laser-cooled so far~\cite{shuman}. A knowledge of the pdm would be useful in determining the long-range dipole-dipole interaction~\cite{Lewin} 
for ultracold molecules in optical lattices. The existence of such an interaction will 
give rise to novel quantum phases, including the elusive supersolid state~\cite{Tapan}. 
Ultracold SrF molecules could also be used for high precision spectroscopy~\cite{Mathavan}~\cite{Steven}. An experiment on 
the search for parity violation using SrF is currently underway~\cite{SH}.

\section{Theory}

The Hamiltonian that we have used is the Dirac-Coulomb Hamiltonian, $H_{DC}$, which is given by:

\begin{eqnarray}
 H_{DC} &=& \sum_i[c\bm{\alpha}.\bm{p_i} + \beta m c^2 - \sum_A \frac{Z_A}{\arrowvert \bm{r_i}-\bm{r_A} \arrowvert}] \nonumber \\
 &+& \sum_{i\neq j} \frac{1}{\arrowvert \bm{r_i}-\bm{r_j} \arrowvert}
\end{eqnarray}

Here, \textit{c} is the speed of light, $\bm{\alpha}$ and $\beta$ refer to the Dirac matrices, and $\bm{p_i}$ refers to the momentum of the $i^{th}$ electron. 
The summation over the electronic coordinates is indicated by i, and that over the nuclear coordinates is indicated by A. $\bm{r_i}$ is the position vector 
from the origin to the site of an electron, and $\bm{r_A}$ is the position vector from the origin to the coordinate of a nucleus. $Z_A$ is 
the atomic number of the $A^{th}$ nucleus.

To obtain the electronic wavefunction, $\arrowvert \psi \rangle$, in our calculations, we use a RCC method. 
The coupled cluster wavefunction can be written as

\begin{eqnarray}
 \arrowvert \psi \rangle = e^T \arrowvert \Phi_0 \rangle
\end{eqnarray}

Here, $\arrowvert \Phi_0 \rangle$ refers to the Dirac-Fock (DF) wavefunction of the ground 
state of the molecule, which is built from single particle four-component spinors. 
$\arrowvert \Phi_0 \rangle$ is taken to be a single determinant 
corresponding to the ionic configuration, Sr: $(5s)^1$ and F: $(2p)^6$, which is an open shell doublet. 
We can visualize the ionic bonding in SrF by considering the difference in the electronegativities of the
constituent atoms. 
T is the cluster operator. In this work, we use the CCSD (Coupled Cluster Singles and Doubles) approximation, where 
$T = T_1 + T_2$. Here, $T_1$ and $T_2$ are the single (S) and double (D) excitation operators respectively. 
They are given by:

\begin{eqnarray}
 T_1&=& \sum_{ap} t_a^p a_p^{\dag} a_a \\
 T_2&=& \sum_{p \neq q, a \neq b} t_{ab}^{pq} a_p^\dag a_q^\dag a_b a_a 
\end{eqnarray}

$t_a^p$ and $t_{ab}^{pq}$ are the cluster amplitudes, where a and b refer to holes and p and q refer to particles. 
$a_p^\dag a_a$ acting on a state means that a hole 'a' is destroyed from that state, and a particle 'p' is created. 
When $a_p^\dag a_a$ acts on a model state, which is the Slater determinant in this case, the resulting state is denoted by 
$\arrowvert \Phi_a^p \rangle$.

The CCSD amplitude equations are:

\begin{eqnarray}
 \langle \Phi_{a}^{p} \arrowvert  e^{-T}H_Ne^{T} \arrowvert \Phi_0 \rangle &=& 0  \\
 \langle \Phi_{ab}^{pq} \arrowvert  e^{-T}H_Ne^{T} \arrowvert \Phi_0 \rangle &=& 0
\end{eqnarray}

The term $e^{-T}H_Ne^{T}$ can be written as $\{H_N e^T\}_C$, using the linked cluster theorem~\cite{Kvasnicka}~\cite{Bishop}. The 
subscript 'C' means that the term is connected. $H_N$ is the normal-ordered Hamiltonian~\cite{Lindgren}.

The important features of our relativistic CCSD method are that it uses the Dirac-Coulomb approximation, it is 
size extensive, and the correlation effects are treated to all orders in the residual Coulomb 
interaction for all possible single and double excitations~\cite{Das}.

In the coupled cluster method, the expectation value of any operator, O, is given by:

\begin{eqnarray}
 \langle O \rangle &=& \frac{\langle \psi \arrowvert O \arrowvert \psi \rangle}{\langle \psi \arrowvert \psi \rangle} \nonumber \\
 &=& \frac{\langle \Phi_0 \arrowvert e^{T \dag} O e^T \arrowvert \Phi_0 \rangle}{\langle \Phi_0 \arrowvert e^{T \dag}e^T \arrowvert \Phi_0 \rangle} \nonumber \\
 &=& \frac{\langle \Phi_0 \arrowvert e^{T \dag} O_N e^T \arrowvert \Phi_0 \rangle}{\langle \Phi_0 \arrowvert e^{T \dag}e^T \arrowvert \Phi_0 \rangle} + \frac{\langle \Phi_0 \arrowvert O \arrowvert \Phi_0 \rangle \langle \Phi_0 \arrowvert e^{T \dag} e^T \arrowvert \Phi_0 \rangle}{\langle \Phi_0 \arrowvert e^{T \dag}e^T \arrowvert \Phi_0 \rangle}
\end{eqnarray}

Here, in the previous line, we have used the fact that $O = O_N + \langle \Phi_0 \arrowvert O \arrowvert \Phi_0 \rangle$~\cite{Crawford}.

We now use the expression~\cite{Cizek}:

\begin{eqnarray}
 \frac{\langle \Phi_0 \arrowvert e^{T \dag} O_N e^T \arrowvert \Phi_0 \rangle}{\langle \Phi_0 \arrowvert e^{T \dag}e^T \arrowvert \Phi_0 \rangle} = \langle \Phi_0 \arrowvert e^{T \dag} O_N e^T \arrowvert \Phi_0 \rangle_C
\end{eqnarray}

Then, the expectation value of the operator O becomes:

\begin{eqnarray}
 \langle O \rangle &=& \langle \Phi_0 \arrowvert e^{T \dag} O_N e^T \arrowvert \Phi_0 \rangle_C + \langle \Phi_0 \arrowvert O \arrowvert \Phi_0 \rangle
\end{eqnarray}

Therefore, pdm of a molecule, d, is given by:

\begin{eqnarray}
 d &=& \frac{\langle \psi \arrowvert D \arrowvert \psi \rangle}{\langle \psi \arrowvert \psi \rangle} \nonumber \\
 &=& \langle \Phi_0 \arrowvert e^{T \dag} D_N e^T \arrowvert \Phi_0 \rangle_C + \langle \Phi_0 \arrowvert D \arrowvert \Phi_0 \rangle \nonumber \\
 &=& \langle \Phi_0 \arrowvert e^{T \dag} D_N e^T \arrowvert \Phi_0 \rangle_C + \langle \Phi_0 \arrowvert (-\sum_i e \bm{r_i} + \sum_A Z_A e \bm{r_A}) \arrowvert \Phi_0 \rangle \nonumber \\
 &=& \langle \Phi_0 \arrowvert e^{T \dag} D_N e^T \arrowvert \Phi_0 \rangle_C + \langle \Phi_0 \arrowvert (-\sum_i e \bm{r_i}) \arrowvert \Phi_0 \rangle \nonumber \\
 &+& \sum_A Z_A e\bm{r_A} \langle \Phi_0 \arrowvert \Phi_0 \rangle \nonumber \\
 &=& \langle \Phi_0 \arrowvert e^{T \dag} D_N e^T \arrowvert \Phi_0 \rangle_C + \langle \Phi_0 \arrowvert (-\sum_i e \bm{r_i}) \arrowvert \Phi_0 \rangle \nonumber \\
 &+& \sum_A Z_A e\bm{r_A}
\end{eqnarray}

where D is the electric dipole moment operator and e is the charge of the electron.   
We have invoked the Born-Oppenheimer approximation in the fourth line of the equations given above. 
SrF is a diatomic molecule, with the fluorine atom chosen to be at the origin. Hence, the pdm can be expressed as

\begin{eqnarray}
 d &=& \langle \Phi_0 \arrowvert e^{T \dag} D_N e^T \arrowvert \Phi_0 \rangle_C \nonumber \\
 &+& \langle \Phi_0 \arrowvert (-\sum_i e \bm{r_i}) \arrowvert \Phi_0 \rangle +  Z_{Sr} e r_e  
\end{eqnarray}

where $r_e$ refers to the equilibrium bond length for SrF. 
The first term captures the electron correlation effects. Note that the leading term in its expansion, namely $\langle \Phi_0 \arrowvert D_N \arrowvert \Phi_0 \rangle$ is zero. 
The second term of the expectation value gives the electronic contribution from the DF calculations. The third part gives the 
nuclear contribution. 
It is important to note that the 
pdm 
depends on the mixing of orbitals of opposite parity. This is naturally achieved in molecules, since the orbitals are hybridized.

\section{Methodology}

The pdm of the ground state of SrF was calculated by combining 
the well known UTCHEM and DIRAC08 codes~\cite{utchem}~\cite{utchemt}~\cite{dirac}.  
The computations for generating the DF orbitals and the transformation of the one and two 
electron integrals from the atomic orbital to the molecular orbital basis~\cite{abe} 
were carried out using the  UTCHEM code. The $C_8$ double group symmetry was used to reduce the computational cost of the calculations~\cite{c8}. 
The DIRAC08 code was used to determine the CCSD amplitudes and the one and two electron integrals were taken from UTCHEM. The electronic part 
of the pdm was then
calculated by using only the linear terms in the coupled cluster wavefunction, since the 
dominant contributions come from them~\cite{Das}. We, hence, evaluate

\begin{eqnarray}
 \langle\Phi_0\arrowvert(1+T_1+T_2)^\dag(-\sum_i e \bm{r_i})_N(1 + T_1 + T_2)\arrowvert\Phi_0\rangle_C \nonumber \\
 + \langle \Phi_0 \arrowvert (-\sum_i e \bm{r_i}) \arrowvert \Phi_0 \rangle
\end{eqnarray}

The nuclear contribution is calculated using the experimental value of the bond length and it is then added to the above quantity to 
obtain the pdm of the molecule.

We performed our calculations with the following choice of uncontracted basis sets: \\
Double zeta (dz) level: Sr:(20s,14p,9d)~\cite{Dyallbasis}~\cite{Sappro} and F: (9s,4p,1d)~\cite{bslF}, \\
Triple zeta (tz) level: Sr:(28s,20p,13d,2f)~\cite{Dyallbasis}~\cite{Sappro} and F: (9s,5p,2d,1f)~\cite{bslF}. \\

The basis sets that we used in our calculations were GTOs (Gaussian Type Orbitals). 
For Sr, we used the exponential parameters taken from the
four-component basis sets obtained by Dyall~\cite{Dyallbasis}. We also added to it diffuse and polarization 
functions from the Sapporo-DKH3 basis sets~\cite{Sappro}. 
We used the exponential parameters of ccpv (correlation consistent-polarized valence) basis sets from the EMSL  Basis Set Exchange Library~\cite{basissetlibrary}~\cite{basissetlibraryt}  for F~\cite{bslF}. 
We have also imposed the kinetic balance condition~\cite{Kenneth}.

\section{Results}

The difference between the CCSD and the DF results gives the correlation contributions. Both these results are given in
Table I for the double and triple zeta basis sets.  

\begin{table}[h] 
 \centering
 \caption{Summary of the calculated results of the present work}
 \label{}
 \begin{tabular}{|r|r|r|r|r|}
 \hline
 $Method$ &$Basis$ &$T_1 dia$ &$pdm (D)$ &$\% frac$ \\
 \hline
  DF&dz &- &2.83 & 18.39 \\
  DF&tz &- &2.95 & 14.93 \\
  CCSD&dz &0.01467 &2.96 & 14.64 \\
  CCSD&tz &0.01743 &3.41 & 1.66 \\
  Expt.&- &- &3.4676(10)  &- \\
 \hline
 \end{tabular}
\end{table}

The values of the pdm have been rounded off to the second decimal place. 
The units are in Debye (D). The experimental bond length of SrF (2.075 Angstrom)~\cite{Herzberg} 
was used in all the calculations. 
The pdm measured from experiment is denoted by 'Expt.' in the table. 
As the size of the basis set increases, the DF values drifts towards the experimental result. 
The results of the CCSD calculations show that the pdm for the triple zeta basis set is closer to the experimental 
value of pdm than that obtained using double zeta basis sets.

Table I also shows the $T_1$ diagnostics (denoted in the table as $'T_1dia'$) for the two runs. They are around 0.01, indicating the 
stability of the single reference calculations for our choice of basis sets. The last column gives the percentage fractional difference between 
the calculation and the experiment.

Table II summarizes the results for the pdm of SrF obtained previously by other methods and experiment.

\begin{table}[h] 
 \centering
 \caption{Summary of the pdms of SrF from previous works and this work}
 \label{}
 \begin{tabular}{|r|r|r|}
 \hline
 $Method$ &$Reference$ &$pdm (D)$  \\
 \hline
  Experiment&Ernst et al ~\cite{Ernstt}&3.4676(10)\\
  Ionic model&Torring et al ~\cite{Torring}&3.67\\
  SCF&Langhoff et al ~\cite{Langhoff}&2.579\\
  CISD&Langhoff et al ~\cite{Langhoff}&2.523\\
  CPF&Langhoff et al ~\cite{Langhoff}&3.199\\
  EPM&J M Mestdagh et al ~\cite{Mestdagh}& 3.6\\
  Ligand field approach&A R Allouche et al ~\cite{Allouche}&3.7875\\
  FD-HF&Kobus etal et al ~\cite{Kobus}&2.5759\\
  CCSD&This work (TZ)&3.41\\
  
 \hline
 \end{tabular}
\end{table}

Torring et al~\cite{Torring} used an ionic model to calculate the pdm of SrF.
Ernst et al experimentally determined the pdm to be 3.4676 (10) D in their work~\cite{Ernstt} a year later. 
The first ab initio calculations were performed by Langhoff et al~\cite{Langhoff}. 
They employed near Hartree-Fock quality Slater basis sets augmented with diffuse and polarization functions, with  
Sr: (12s,10p,7d,3f) and F: (6s,5p,4d,2f). They calculated the 
pdm of the molecule using 
the single reference CISD (Configuration Interaction Singles and Doubles) and 
the CPF (Coupled Pair Functional) approaches. 
The former gave a value of 2.523 D, and the latter 3.199 D, and both these results used
the experimental value of the bond length (2.075 Angstrom). 
CISD is not size extensive, and for a given level of excitation, CCSD includes more correlation effects than CISD. This may have led to certain
missing terms that could have played a crucial role in cancellations with other terms in the expectation value. 
The CPF approach is size consistent, and the result is in better agreement with the experimental value. However, 
SrF is large enough for relativistic effects to be pronounced, but in this paper, the treatment 
is non-relativistic.  
Mestdagh and Vistocot~\cite{Mestdagh} used an electrostatic polarization model (EPM). They used the experimental value of bond length for 
their calculations, and obtained a value of 3.6 D. 
Allouche et al~\cite{Allouche} used a ligand field approach for their calculations. The largest basis sets they used for Sr had 10 radial functions. 
The work by Kobus et al~\cite{Kobus} was to compare the dipole moments obtained from finite basis set with finite difference Hartree-Fock calculations. 
We have mentioned the work for completeness. 
 Our fully relativistic calculations gave a result of 3.41 D with Sr: (28s,20p,13d,2f) and F: (9s,5p,2d,1f) basis sets (at the tz level).

We now proceed to focus on the contributions from the individual terms in the pdm given in equation 12. 
We consider the case for the largest basis set that we have used in our calculations, since the SrF wavefunction built from 
it is the closest to the actual wavefunction among the basis sets that we have chosen. 
We only deal with the electronic contribution to the pdm, because as dicussed earlier, the nuclear part is evaluated separately.  
 The ket can be written in terms of either the DF, single-excitation or double-excitation  
determinants. The bra can also be expressed similarly. The resulting contributions to the pdm from the nine 
terms are given below.

\begin{table}[h] 
 \centering
 \caption{Contributions from the individual terms to the pdm}
 \label{}
 \begin{tabular}{|r|r|}
 \hline
  Term &Contribution \\
 \hline
  DF &-375.81 \\
  D$T_1$ &0.26 \\
  D$T_2$ &0.00 \\
  $T_1^\dag D$ &0.26 \\
  $T_1^{\dag} D T_1$ &-0.04 \\
  $T_1^{\dag} D T_2$ &0.02 \\
  $T_2^{\dag} D$ &0.00 \\
  $T_2^{\dag} D T_1$ &0.02 \\
  $T_2^{\dag} D T_2$ &-0.06 \\

 \hline
 \end{tabular}
\end{table}

In Table III, the first column specifies the different terms in equation 12, while the second column gives 
their respective values (rounded off to two decimal places). 
The values for the electronic terms are all in Debye. They add up to -375.35 D. The nuclear contribution is 
378.76 D. The resulting pdm is, hence, 3.41 D . 
The electronic contribution from the DF calculations is -375.81 D, and 
hence, the final value for the pdm at the DF level is 2.95 D. CCSD accounts for the correlations that make up 
the remaining 0.46 D, which is about 13.5 percentage of the final value of the pdm.

We can see that the dominant contribution is from the DF term. The matrix elements of the pdm operator between 
the DF and single excitation determinants (and their Hermitian conjugate), which incorporate an important class 
of pair correlation effects~\cite{Natarajhs}, make the
largest contribution among the CCSD terms. These correspond to the $D T_1$ and the $T_1^\dag D$ terms.

We give in Figure 1 the Goldstone diagrams that correspond to the terms in equation 12.

\begin{figure}[h]
   \centering
\psfig{file=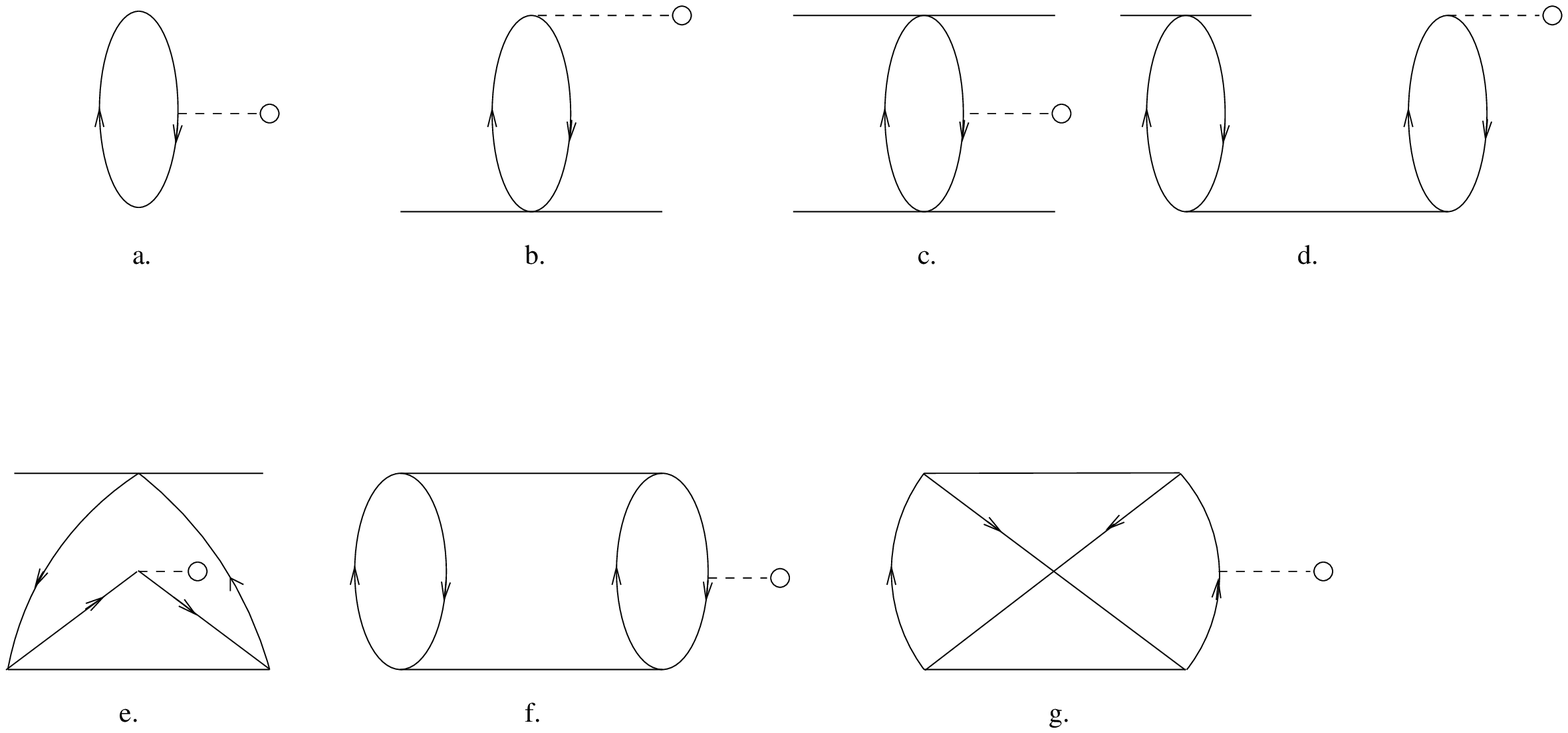,height=2.2in,width=3.2in,angle=0}
\caption{Goldstone diagrams for pdm: a. DF term, b. $D T_1$, c. $T_1^{\dag} D T_1$, d. and e.direct and exchange diagrams 
respectively of the $T_1^{\dag} D T_2$ term, f. and g. direct and exchange diagrams 
respectively of the $T_2^{\dag} D T_2$ term.}
\label{figure:Figure 1}
\end{figure}

The possible sources of error could arise from not considering higher order correlation effects and the choice of basis sets. 
Apart from the DF term, the $OT_1$ term and its conjugate, given in Table 3, the rest of the terms add to -0.06 D. 
It is reasonable to assume that 
the error from higher order correlations does not exceed this value. We can approximately state that this error is about $\pm$0.1 D. 
Taking into account the difference between TZ and QZ results that we obtained for the pdm in each of the three choices of 
basis sets, namely, Dyall, Sapporo and a 
combination of Dyall and Sapporo sets, 
we could say that a conservative estimate of the difference between these results is 2 percentage, which is 
around 0.1 D. 
A ballpark estimate of the error would, hence, be around $\pm$0.2 D for 
the pdm of SrF. 

\section{Conclusion}

In order to test the accuracy of our RCC method, we have calculated the pdm 
of SrF, using 
two kinds of uncontracted Gaussian basis sets, based on the Dyall plus Sapporro (for diffuse and polarization functions) and ccpv basis sets.
Our result for the larger of the two basis sets; i.e. the tz basis with Sr: (28s,20p,13d,2f) and F: (9s,5p,2d,1f), gave 
a value that is closer to that of experiment than that of the other basis set.   
The largest contribution to pdm  arises from the DF term, and
next in importance are the single excitations which embody an important type of pair correlation
effect.

Our result for SrF pdm with a high precision measurement of this quantity is in better agreement than those of all 
the other earlier calculations, and it is a little over one and a half percentage of the measured value. 
With the error in the calculation estimated to be $\pm$0.2 D, the experimental value falls within 
our estimated error range. 
This is a testament to the 
power of the RCC method to account for relativistic and  
electron correlation effects.  
Our present work sets the stage for exploring many other properties of molecules using 
the RCC method.

\section{Acknowledgment}

The computational results reported in this work 
were performed on the high performance computing facilities of Indian Institute of Astrophysics (IIA), Bangalore, 
on the Hydra and Kaspar clusters. 
We thank Prof. Hiroshi Tatewaki for useful discussions on basis functions. 
VSP would like to acknowledge 
the help of Anish Parwage, Engineer C, IIA, India, for his help with installing codes on the clusters. 
The DiRef database was extremely useful in 
searching for literature~\cite{diref}. The research was supported by JST, CREST. MA thanks MEXT for financial support.

\end{document}